\documentclass[12pt,a4paper]{article}

\usepackage{authblk}
\usepackage{graphicx} 
\usepackage{lineno}
\usepackage{amsmath}
\usepackage{url}
\usepackage{cite}
\usepackage{tikz-feynman}
\usepackage{caption}
\usepackage{subcaption}
\usepackage{slashed}
\usepackage{array}
\usepackage{comment}
\usepackage{ulem}

\graphicspath{ {Images/} }

\newcommand\vectorDMdiag{
\begin{tikzpicture}[baseline=-0.09cm]
        \begin{feynman}
            \vertex[dot] (a) {};
            \vertex[dot,right= of a] (b) {};
            \vertex(c) at (-1.5, 1)  {$\bar{q}$};
            \vertex(c2) at (-0.75, 0.5)  {};
            \vertex(c3) at (0.4, 1.4)  {};
            \vertex(d)  at (-1.5, -1) {$q$};            
            \vertex[above right= of b] (e) {\textcolor{gray}{DM}};
            \vertex[below right= of b] (f) {\textcolor{gray}{DM}};

            \diagram* {
                (a) -- [fermion,thick] (c),
                (c2) -- [gluon] (c3),
                (d) -- [fermion,thick] (a),
                (a) -- [photon, edge label=$Z'$] (b),
                (b) -- [gray,fermion,thick] (e),
                (b) -- [gray,fermion,thick] (f),
            };          
        \end{feynman}        
\end{tikzpicture}
}
\newcommand\pseudoDMdiag{
\begin{tikzpicture}[baseline=-0.09cm]
                \begin{feynman}
                    \vertex[dot] (a) {};
                    \vertex[dot,left= of a] (b) {};
                    \vertex[above right= of a] (c) {\textcolor{gray}{DM}};
                    \vertex[below right= of a] (d) {\textcolor{gray}{DM}};
                    \vertex[dot,above left=20pt of b] (e) {};
                    \vertex[dot,below left=20pt of b] (f) {};
                    \vertex(g) at (-3.5, 1) {$g$};
                    \vertex(g2) at (-2.75, 0.75) {};
                    \vertex(g3) at (-1, 1.4)  {};
                    \vertex(h) at (-3.5, -1)  {$g$};
                    \diagram* {
                        (a) -- [fermion,gray] (c),
                        (d) -- [fermion,gray] (a),
                        (g2) -- [gluon] (g3),
                        (a) -- [scalar,edge label'=$\phi$] (b),
                        (b) -- [edge label'=$t$] (e),
                        (b) -- [edge label=$t$] (f),
                        (e) -- [gluon,thick] (g),
                        (h) -- [gluon,thick] (f),
                        (f) -- [thick,edge label=$t$]  (e) 
                    };          
                \end{feynman}        
        \end{tikzpicture}
}

\newcommand\squarkDMdiag{
\begin{tikzpicture}[baseline=-1cm]
        \begin{feynman}
            \vertex[dot] (a) {};
            \vertex[dot, below=30pt of a] (b) {};
            \vertex[left= of a] (g1) {$q$};
            \vertex[right=72.5pt of a] (t) {\textcolor{gray}{DM}};
            \vertex[left= of b] (g2) {$g$};
            \vertex[dot,right=30pt of b] (tb) {};
            \vertex[right= of tb] (tb2) {\textcolor{gray}{DM}};
            \vertex[below right=40pt of tb] (tb3) {$q$};
            \diagram* {
                (g1) -- [fermion,thick] (a),
                (g2) -- [gluon,thick] (b),
                (a) -- [fermion,gray] (t),
                (tb) -- [fermion,gray] (tb2),
                (tb) -- [fermion] (tb3),
                (tb) -- [scalar,edge label=$\tilde q$] (b),
                (b) -- [scalar,edge label=$\tilde q$] (a),
            };          
        \end{feynman}        
\end{tikzpicture}
}

\title{Universal New Physics Latent Space}
\author[1]{Anna Hallin}
\author[1]{Gregor Kasieczka}
\author[2]{Sabine Kraml}
\author[3]{Andr\'e Lessa}
\author[1]{Louis Moureaux}
\author[1]{Tore von Schwartz}
\author[4]{David Shih}

\affil[1]{Institut f\"ur Experimentalphysik, Universit\"at Hamburg, Luruper Chaussee 149, 22761 Hamburg, Germany}
\affil[2]{Laboratoire de Physique Subatomique et de Cosmologie (LPSC), Universit\'e
Grenoble-Alpes, CNRS/IN2P3, F-38026 Grenoble, France}
\affil[3]{Centro de Ci\^encias Naturais e Humanas, Universidade Federal do ABC, Santo Andr\'e, 09210-580 SP, Brazil}
\affil[4]{NHETC, Department of Physics and Astronomy, Rutgers University, Piscataway, NJ 08854, USA}

\date{\today}

\begin{document}

\maketitle

\begin{abstract}
We develop a machine learning method for mapping data originating from both Standard Model processes and various theories beyond the Standard Model into a unified representation (latent) space while conserving information about the relationship between the underlying theories.
We apply our method to three examples of new physics at the LHC of increasing complexity, showing that models can be clustered according to their LHC phenomenology: 
different models are mapped to distinct regions in latent space, while indistinguishable models are mapped to the same region. 
This opens interesting new avenues on several fronts, such as
model discrimination, selection of representative benchmark scenarios, and identifying gaps in the coverage of model space.
\end{abstract}

\clearpage

\section{Introduction}
With Run 3 of Large Hadron Collider (LHC) underway, the roadmap for the High-Luminosity LHC established, and discussions of future colliders ongoing, searches for new physics are at a crossroads. The simplest and/or most popular extensions of the Standard Model (SM) are severely challenged by the null results of previous LHC runs.\footnote{For comprehensive reviews of Run~2 searches, see \cite{ATLAS:2024lda,CMS:2024zqs,CMS:2024bni}.} 
Looking beyond these models quickly opens up a vast theory space of all kinds of possible Beyond the Standard Model (BSM) scenarios~\cite{Bose:2022obr,Fox:2022tzz}. This large theory space makes it difficult to decide which signatures should be probed next with high priority. Optimizing analyses for specific target models needs to be balanced against a broad coverage of signature space and convenient re-interpretability in terms of alternative theoretical scenarios~\cite{Bailey:2022tdz}.

The large theory space also makes mapping the implications of theoretical models to the LHC data a non-trivial and computationally intensive effort. It is unavoidable that different models will yield predictions that are very similar within the current experimental capabilities. Conversely, a single experimental result will have implications for the feasibility of multiple models, not all of which are explicitly considered by the experiment. Modern anomaly detection based search strategies
\cite{Kasieczka:2021xcg,Aarrestad:2021oeb,Karagiorgi:2022qnh,Belis:2023mqs} complicate the matter further by aiming for model-agnostic sensitivity to various outliers and overdensities in the data. While this wider sensitivity might be useful in accelerating the discovery of BSM physics, it makes the assessment of which models were potentially covered by a given search more difficult.

In this work, we propose a novel approach using 
machine learning (ML) to address the issues of re-interpretation and experimental coverage of possible BSM signatures. 
Using a fully connected encoder architecture, we build an abstract low-dimensional representation of BSM models (a so-called latent space) so that experimentally indistinguishable models are mapped to the same region in this space, while models which are far apart in latent space have distinct phenomenologies. 
Once realized, such a universal new physics latent space would simplify the re-casting of results by projecting into the latent space as well as identifying insufficiently explored regions. Models with similar signal properties would not have to be individually searched for but could be replaced by a single representative scenario, provided appropriate reinterpretation is made possible~\cite{LHCReinterpretationForum:2020xtr,Bailey:2022tdz} 
and that the choice of representative scenarios does not leave significant holes in latent space.
In addition, since distances between two models in latent space give an approximate measure of how experimentally distinct these models are, the latent space can also be useful for tackling the inverse problem, i.e. inferring from data which kind of BSM scenario is realised.

Several ML techniques have been suggested for the task of determining what BSM parameters are allowed within the constraints of the data. For a review of these, see~\cite{Baruah:2024gwy}. More specifically related to our work, the connection between the latent space and the underlying physics has previously been studied in the context of strings~\cite{Mutter:2018sra}, $W$-jets~\cite{Collins:2021pld}, supersymmetry (SUSY)~\cite{Baretz:2023mra} and QCD vs.\ BSM jets~\cite{Park:2022zov} where the aims and strategies of the latter are the most similar to ours. But while~\cite{Park:2022zov} deals with single jets and constructs a loss function that aims to preserve the distance between two events (defined via metrics on the data space and the latent space), our model operates on event level rather than jet level and uses a contrastive loss. Some other attempts at constructing mappings between theory space and observables can be found in Refs.\cite{Arkani-Hamed:2005qjb,Kane:2007pp,Heckman:2009bi,Bornhauser:2012iy,Caron:2016hib,Komiske:2020qhg}.

The rest of this paper is organised as follows: section~2 will introduce the machine learning approach; sections~3--5 then apply this technique to three different datasets with increasing complexity: simulated gluino decays from 
a gluino-neutralino simplified model, simplified Dark Matter (DM) models leading to one hard jet and missing energy, and the public Dark Machines Anomaly Challenge dataset; section~6 offers a summary and conclusions. 
Additional plots for the three datasets are collected in the Appendix.

\section{Method}
\label{sec:method}

In this work, we learn a mapping of the high dimensional theory space of BSM theories to a lower dimensional latent space using an encoder network trained on physical (detector-level) observables such as missing transverse energy (MET) and jet variables. 

Since a set of events better represents the signal features than a single event, a single input data point ($\vec{X}_i$) for the network corresponds to the concatenation of observables (or input features) for a set of 10 Monte Carlo (MC) events generated by the same BSM model\footnote{By BSM model we mean a specific BSM scenario with fixed model parameters. For instance, for the simplified gluino-neutralino model in the next section, two distinct sets of gluino and neutralino masses are considered as two different BSM models.}, 
e.g.:
\begin{equation}
    \vec{X}_i = \left(\left\{{\rm MET}, p_T(j_1),...\right\}_{\rm event 1},\;...,\;\left\{{\rm MET}, p_T(j_1),...\right\}_{\rm event 10} \right)
\label{Xi}
\end{equation}
The events for each set were chosen at random, and the sets were recreated for each epoch.

The encoder network architecture is a fully-connected MLP with hidden layer dimensions (384, 512, 256, 128, \dots, 4) and ReLU activations. The final hidden layer outputs to the latent space. We set the latent space dimension to two for ease of visualization. Increasing the latent space to higher dimension was not found to provide any further benefit in the different scenarios studied in this work.

We use a contrastive loss function which is designed to learn the essential differences between the various models. Contrastive learning \cite{Hadsell1640964} relies on determining whether two data points are similar or not. Based on this similarity measure, the data points either get separated or pulled together in latent space. The resulting mapping should encode the pairwise relationships between the data points.

Concretely, each pair of event sets $\vec{X}_i$ and $\vec{X}_j$ is assigned a label $Y_{ij}$ that is set to zero for points from the same model (similar points) or to 1 for points from distinct models (dissimilar points).  
In other applications, like image recognition~\cite{Chen:2020pse} or natural language processing~\cite{cheng2023improving}, similar samples are often generated using data augmentation~\cite{khosla2021supervised}, which makes it possible to use this method even without labels. In this work, we instead use the underlying model to define the similarity. The contrastive loss function has the form
\begin{equation}
    \mathcal{L}(\theta) = \sum_{i\ne j} L(\theta, (Y_{ij}, \vec{X}_i, \vec{X}_j)) \,,
\end{equation}
where 
\begin{equation}
    L(\theta, (Y_{ij}, \vec{X}_i, \vec{X}_j)) = (1-Y_{ij})\frac{1}{2} D^2_\theta + Y_{ij}\frac{1}{2}\left(\max\left(0, d -D_\theta\right)\right)^2 \,.
\label{L_CL}
\end{equation}
Here, $D_\theta\equiv D_\theta(\vec{X}_i, \vec{X}_j)$ is the Euclidian distance in the latent space, $\theta$ are the weights of the encoder network being trained and $d$ is an additional hyperparameter that is used to deal with unbounded representation spaces. Note that for bounded representation spaces it is not necessary to introduce a maximum distance. In our case, however, we do not impose any bounds on the latent space and $d$ is needed to ensure a converging training.
In what follows, we set $d=1$.

With the above definition, the loss function is minimized if $D_\theta^2$ is small for similar points and large for dissimilar points. If different models occupy disjoint (i.e.\ non-overlapping) regions of phase space, the loss can be trivially set to zero by collapsing each model to a distinct point in latent space, separated from all the other points by distances $D_\theta\ge 1$. The interesting, non-trivial action of the loss arises when different models occupy overlapping regions of phase space: when inputs from these regions of phase space are fed to the network, the loss tries to minimize $D_\theta$ when the inputs come from the same model, but tries to maximize $D_\theta$ when the inputs come from different models. This tension results in a nonzero value of $D_\theta$, i.e.\ the models cannot be collapsed to a single point by the network. Instead, they are mapped to non-trivial regions in the latent space which characterize and represent the phase space regions occupied by each model.

Three datasets, described in more detail below,  were tested with this contrastive learning model. They all feature jets plus MET signatures but represent different levels of complexity for model discrimination. The input data points $\vec{X}_i$ for the training (see Eq.~\eqref{Xi}) were thus made of the MET plus various kinematic variables of the jets ($p_T$, $\eta$, $\Delta\phi$, etc.); the specific inputs for each dataset will be described below. Every input feature was normalized to have mean zero and standard deviation one. The training was done using Adam~\cite{kingma2017adam} as optimizer with a learning rate of $10^{-5}$. Once trained, the model state with the lowest validation loss was selected.

\section{MSSM gluino simplified model}
\label{sec:susy_dataset}

\subsection{Dataset}

\begin{figure}[th]
\centering
\includegraphics[width=5.6cm]{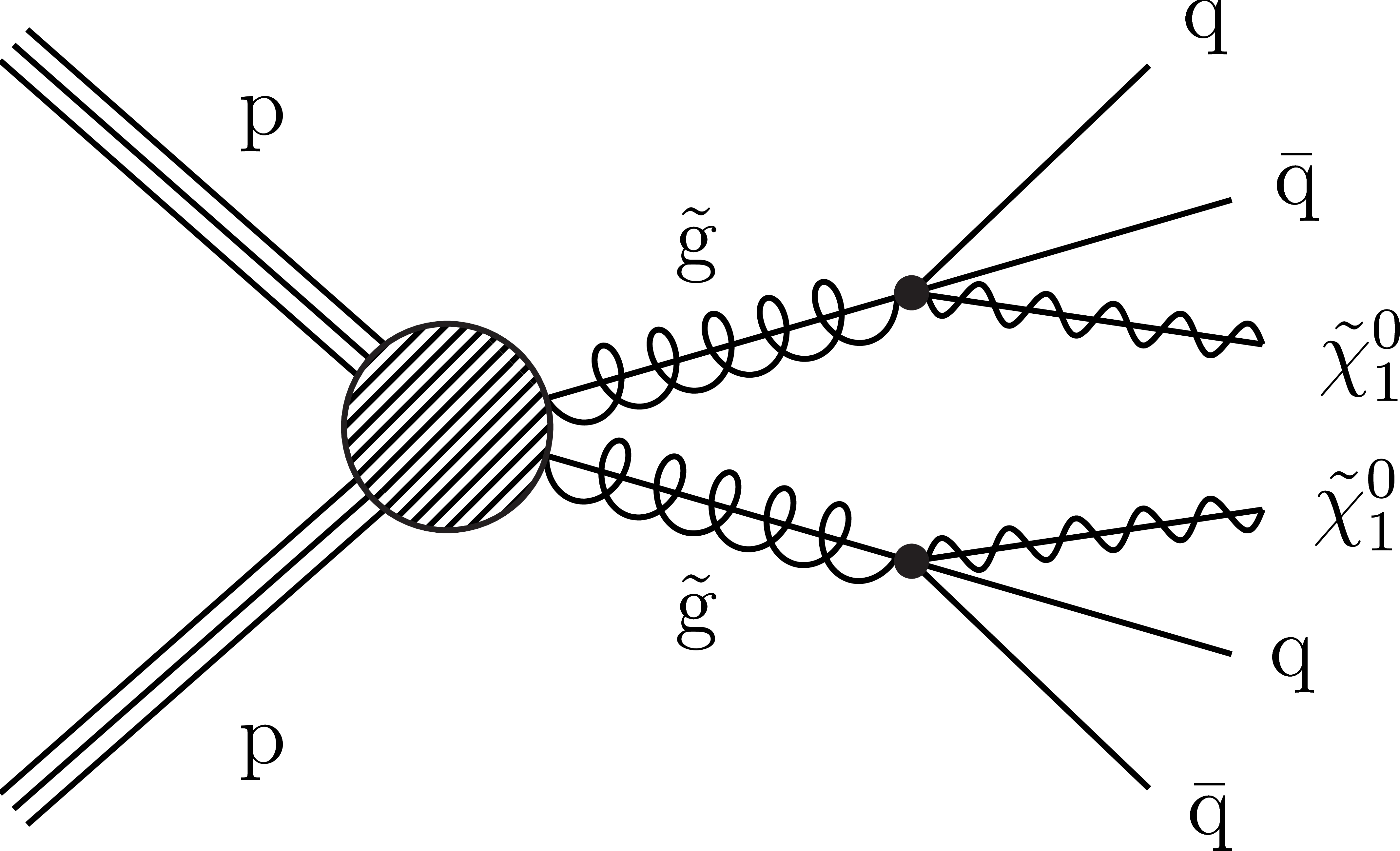}
\caption{Process simulated for the MSSM gluino dataset.
}
\label{fig:feynman_susy}
\end{figure}

The first dataset to be studied is a supersymmetric scenario, with pair-production of gluinos in the Minimal Supersymmetric Standard Model (MSSM). Each gluino is assumed to decay with 100\% branching ratio to a neutralino $\tilde\chi^0_1$ and a quark-antiquark pair, $\tilde g\to q\bar q\tilde\chi^0_1$, as illustrated in Fig.~\ref{fig:feynman_susy}. This leads to events with four hard jets and large MET. The dataset was generated with MadGraph5\_aMC@NLO\,v3.3.0~\cite{Alwall:2014hca}, Pythia\,8.306~\cite{10.21468/SciPostPhysCodeb.8,10.21468/SciPostPhysCodeb.8-r8.3}, Delphes\,3~\cite{deFavereau:2013fsa}, and NNPDF\,2.3~\cite{Ball:2012cx} at LO. Nine distinct BSM models were simulated, with gluino masses of $m_{\tilde g}=1.1$, 1.6 and 2.1~TeV and neutralino masses of $m_{\tilde\chi^0_1}=0.1$, 0.5 and 0.9~TeV. The events from all mass configurations were combined into one single dataset.

The features used in training from this dataset are the $p_T, \eta, m$ and $\Delta\phi$ of the first four jets in the event, where $p_T$ is the transverse momentum, $\eta$ the pseudorapidity, and $m$ the invariant mass of the jet; $\Delta\phi$ is the difference in azimuthal angle between the jet and the MET, defined by
\begin{equation}
    \Delta\phi = \min \left(\left|\phi_j-\phi_{\rm MET} \right|, 2\pi- \left|\phi_j-\phi_{\rm MET} \right|\right) .
    \label{eq:delta_phi}
\end{equation}
In the set of features we also include the MET and the invariant mass of all possible jet pairs:
\begin{equation}
    m^2_{j_a,j_b} = 2\,p_{T_a}\,p_{T_b} \left[\cosh(\eta_a - \eta_b) - \cos(\phi_a - \phi_b)\right] .
\label{eq:MSSM_pairwise}
\end{equation} 

The distributions of the leading jet $p_T$, the MET and $m^2_{1,2}$ are shown in Fig.~\ref{fig:histogram_3jets_susy}, while the full set of features for all four jets can be found in Appendix~\ref{app:MSSM}.
Since the amount of energy going into jets depends mostly on the gluino-neutralino mass difference ($\Delta m$), we expect harder jets and a harder MET spectrum for larger $\Delta m$. This is confirmed in Fig.~\ref{fig:histogram_3jets_susy}, where the hardest MET spectrum corresponds to $\Delta m = 2$~TeV and the softest to $\Delta m = 0.2$ TeV. A similar behavior can be observed for the $p_T$ and the invariant mass $m_{1,2}^2$. Furthermore, we see that models with similar values of $\Delta m$ have similar features and are harder to distinguish experimentally.\footnote{Note that we are not making use of the total rate of events, which depends on the gluino cross-section and could in principle be used to distinguish between models with different gluino masses.}

\begin{figure}
    \centering
    \includegraphics[width=\textwidth]{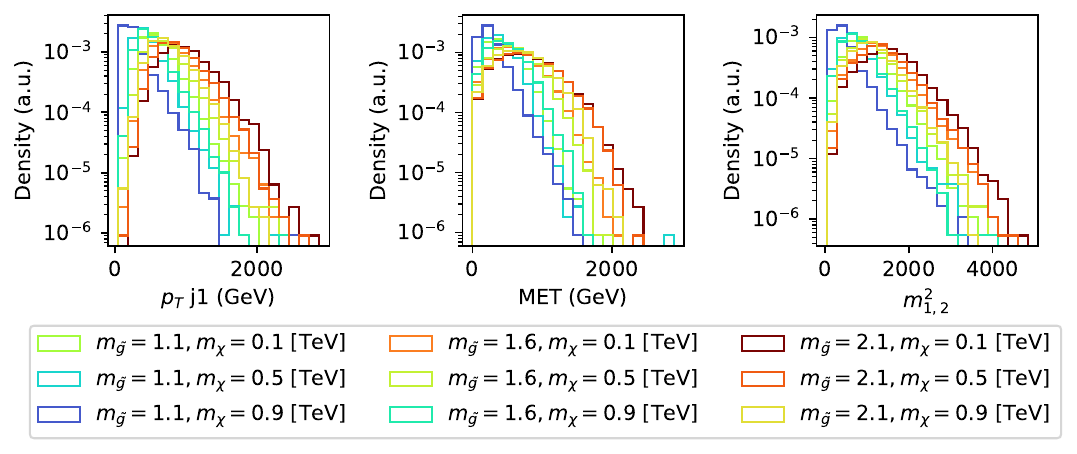}
    \caption{Distributions of the leading jet $p_T$, plus MET and $m^2_{1,2}$, for each of the mass configurations in the MSSM gluino dataset. }
    \label{fig:histogram_3jets_susy}
\end{figure}

\subsection{Results}
\label{sec:SUSY_results}

\begin{figure}[t!]
\centering
\includegraphics[width=0.85\textwidth]{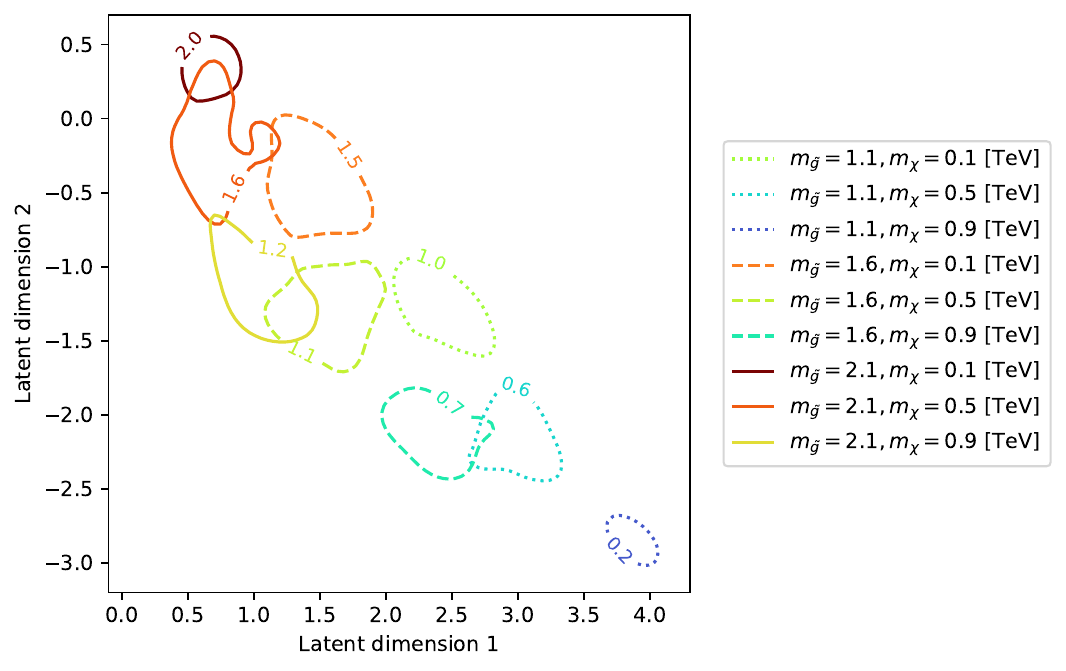}
\caption{Resulting latent space for the MSSM gluino dataset, showing contours corresponding to a CDF value of 0.5 for each model. The different contours are colored based on the mass difference $\Delta m=m_{\tilde g}-m_{\tilde\chi^0_1}$, which is also given as the contour label (in TeV); dark red: $\Delta m=2$~TeV; red: $\Delta m=1.6$ (dashed) and $1.5$ (solid)~TeV; green: $\Delta m=1.2$ (dash-dot), $1.1$ (dashed) and $1.0$ (solid)~TeV; cyan: $\Delta m=0.7$ (dash-dot) and $0.6$ (dashed)~TeV; blue: $\Delta m=0.2$~TeV. The linestyles are chosen based on the gluino mass: solid for $m_{\tilde g}=2.1$~TeV, dashed for $m_{\tilde g}=1.6$~TeV, and dotted for $m_{\tilde g}=1.1$~TeV.}
\label{CL_gluino_dataset}
\end{figure}

Training the encoder with data from the nine models and using the signal features discussed in the previous section, we obtain the latent space mapping shown in Fig.~\ref{CL_gluino_dataset}.
The plot was created by first mapping the data to the embedding space and then fitting a probability density in this space using a kernel density estimation with a Gaussian kernel. The contours in the plot represent 
equal-density contours that enclose 50\% of the probability density, corresponding to a cumulative distribution function (CDF) value of 0.5. This means that half of the events fall within these contours. The colors of the different contours correspond to the mass difference $\Delta m$ between the gluino and the neutralino. We see that the network has not only managed to separate the different datasets into different clusters, but it has also ordered the latent space according to $\Delta m$ along the diagonal axis. This behavior agrees with the strong correlation that the signal features have with $\Delta m$, as shown in Fig.~\ref{fig:histogram_3jets_susy}. Furthermore, somewhat orthogonal to this main diagonal, the network seems to sort the models according to $m_{\tilde g}$, with smaller gluino masses corresponding to larger values for the $x$-axis. This separation is not trivial, given that the signal features are weakly dependent on the gluino mass.

\section{Dark Matter simplified model}
\label{sec:dark_matter_dataset}
\subsection{Dataset}

\begin{figure}[!t]
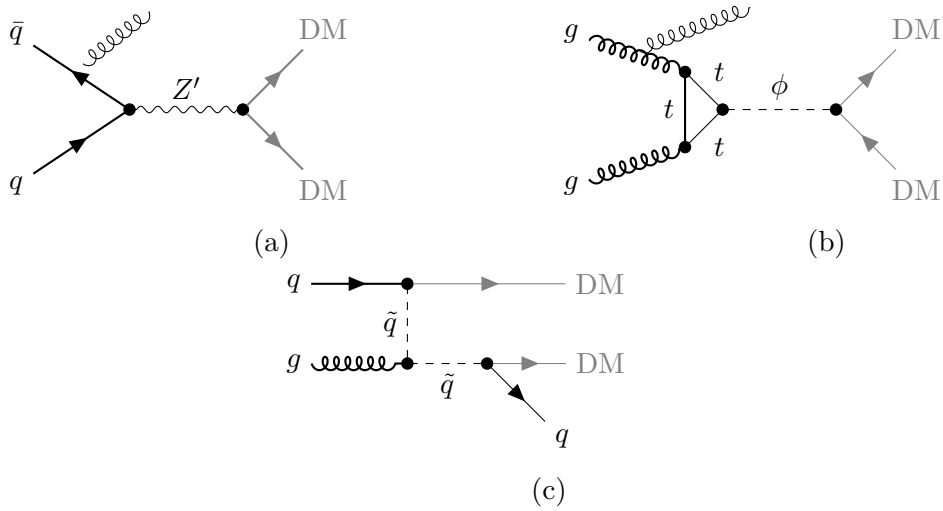

\centering
\begin{subfigure}[b]{0.45\textwidth}
{\small \vectorDMdiag} \label{fig:vectorDM}
\caption{}
\end{subfigure}
\begin{subfigure}[b]{0.45\textwidth}
{\small \pseudoDMdiag} \label{fig:pseudoscalarDM}
\caption{}
\end{subfigure}
\begin{subfigure}[b]{0.45\textwidth}
{\small \squarkDMdiag} \label{fig:squarkDM}
\caption{}
\end{subfigure}
\caption{Examples of diagrams for DM production in the: a)~vector mediator, b)~pseudoscalar mediator and c)~squark mediator models. \label{fig:DMdiagrams}}
\end{figure}

In the gluino-neutralino scenario discussed above, the theory space of BSM models was two dimensional ($m_{\tilde g},m_{\tilde \chi_1^0}$) and was successfully mapped to a two dimensional latent space. As the next step in complexity, we analyze distinct classes of two dimensional models, which corresponds to adding a third (discrete) dimension to the theory space.
To this end, we consider DM production at the LHC, which is typically searched for in events with a single hard jet plus missing energy (additional soft jets are also present from initial and final state radiation, ISR and FSR). The following three models are used:
\begin{itemize}
    \item Vector mediator: in this model, the DM candidate is pair-produced through an s-channel spin-1 mediator, as shown in Fig.~\ref{fig:DMdiagrams}a.
    \item Pseudoscalar mediator: this model is similar to the above, but with the production mediated by a pseudoscalar, which is produced through the gluon fusion diagram shown in Fig.~\ref{fig:DMdiagrams}b.
    \item Squark mediator: unlike the previous cases, this model presents a DM-squark associated production, where the squark further decays to DM and a hard jet, as shown in Fig.~\ref{fig:DMdiagrams}c.
\end{itemize}
In all these models, the DM candidate is a Majorana fermion and a hard jet is present either from ISR in the vector and pseudoscalar mediator models, or from the squark decay. We also point out that in the three scenarios the DM production is dominated by distinct initial-state partons: the vector mediator is produced from $q \bar{q}$ collision, the pseudoscalar from $g g$ collision, and  the squark-DM production requires a $q g$ initial state. 
Monte Carlo events were generated for the following ranges of DM and mediator masses:
\begin{align*}
    m_{\rm DM} & = 100-900 \mbox{ GeV in steps of 100 GeV} \,,\\
    M_{\rm med} & = 600-2000 \mbox{ GeV in steps of 200 GeV} \,,
\end{align*}
where $M_{\rm med} = M_{Z'},M_{\phi},M_{\tilde q}$ for the vector, pseudoscalar and squark models, respectively and $m_{\rm DM} \leq 2 M_{\rm med}$ was imposed.
For each mass combination, 50k MC events were generated at LO using {\tt MadGraph5\_aMC@NLO}~v.3.4.1\cite{Madgraph1,Madgraph2}, {\sc Pythia 8}~v8.307~\cite{10.21468/SciPostPhysCodeb.8} and Delphes v3.5.0~\cite{deFavereau:2013fsa} with the PDF set {\tt NNPDF31\_nnlo\_as\_0118} and the default Delphes card for the CMS detector. In the case of the vector and pseudoscalar mediator models, the DM production was computed with one additional extra jet and the MLM matching scheme was used with {\tt xqcut} $= M_{Z'}/15$ or $M_{\phi}/12$ in order to properly model ISR jets.

\begin{figure}
    \centering
    \includegraphics[scale=0.6]{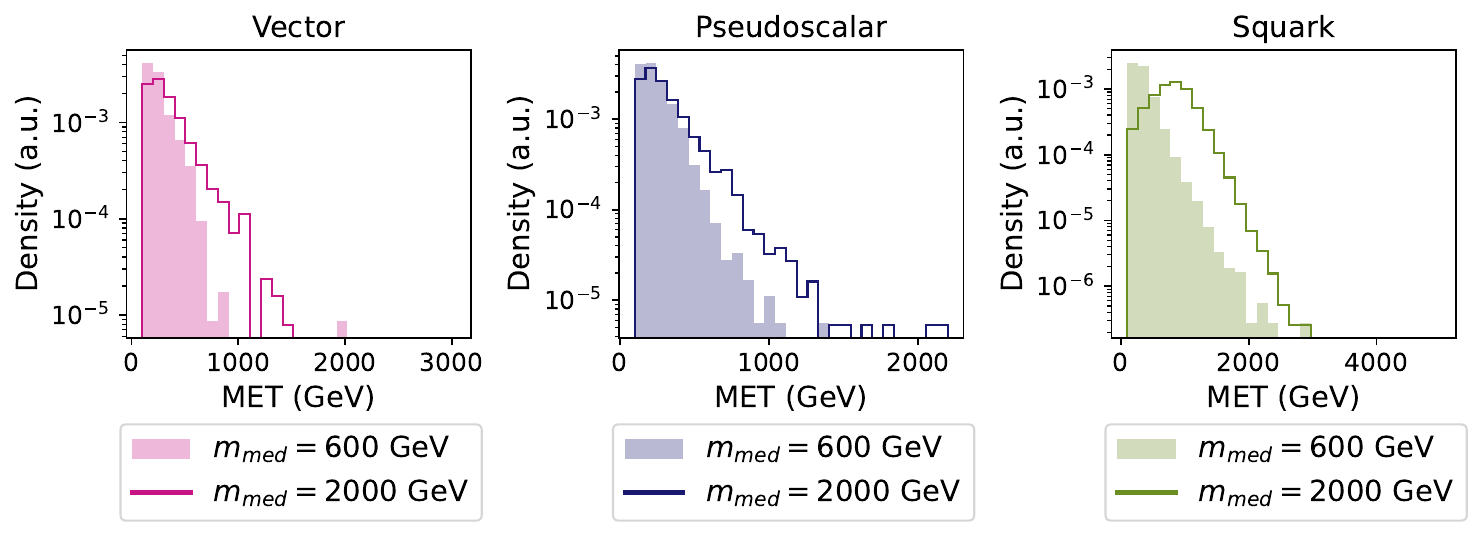}
    \caption{MET distribution of the three different mediator types with $m_{\rm DM}=100$~GeV and the highest and lowest of the mediator masses. It is clearly visible that while the peak of the distribution stays in the same place for the vector and pseudoscalar mediators, for the squark mediator it gets shifted with higher mediator mass.}
    \label{fig:alltypes_MET_extremes}
\end{figure}

The $p_T$, $\eta$, $\Delta\phi$ and mass of the leading two jets and the MET of the event are used as input features, with $\Delta\phi$ as defined in Eq.~\eqref{eq:delta_phi}. In order to satisfy minimum trigger requirements, a cut on the transverse momentum is applied by demanding $p_{T,1} > 100$~GeV and $p_{T,2} > 50$~GeV. 
Since MET is expected to be the leading feature for these models, we show in Fig.~\ref{fig:alltypes_MET_extremes}  the MET distribution for the highest and lowest mediator masses for the scenario where $m_{\rm DM} = 100$~GeV.
The full feature distributions of the three mediator scenarios are shown in Appendix~\ref{sec:appendix_dark_matter} for the different mediator masses. 
We see that the vector and pseudoscalar models have a MET distribution which shifts slightly with the increase of the mediator mass. On the other hand, the squark model presents a distribution which is more strongly correlated to the mediator mass. This is expected since the $p_T$ of the hard jet produced by the squark decay directly depends on the squark-neutralino mass difference. It is also important to notice that, since the scalar and vector mediators are produced on-shell, the signal features have a weak dependence on $m_{\rm DM}$.

\subsection{Results}
\label{sec:dark_matter_results}

\begin{figure}[t!]\centering
\includegraphics[width=\textwidth]{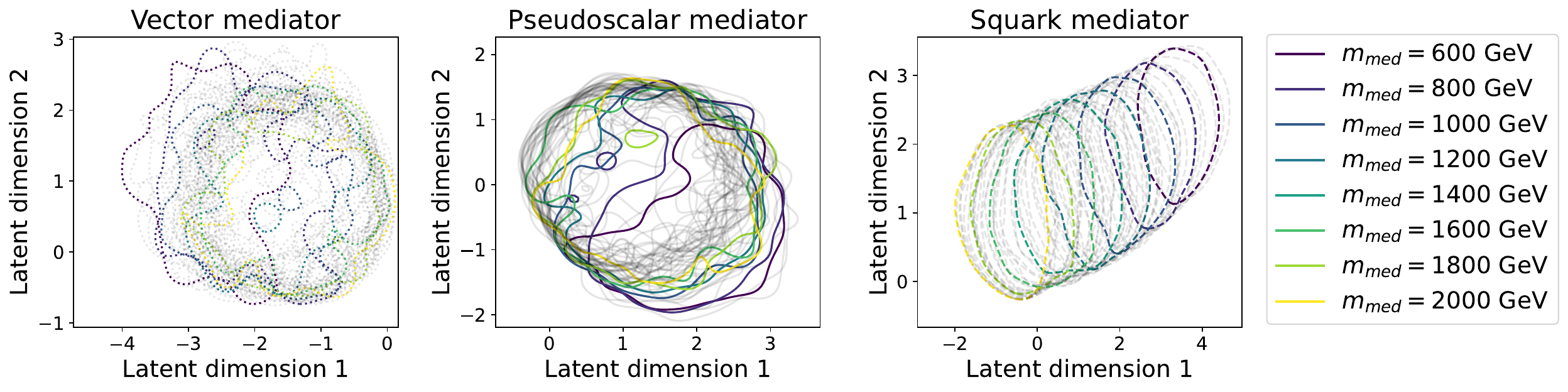}
\caption{Latent space for the separate embeddings of the different mediator types. The contours correspond to a CDF value of 0.5 for varying mediator masses, with $m_{\rm DM}=100$~GeV. The contour lines for all other mass configurations are drawn in light gray.}
\label{mediator_single_comparison}
\end{figure}

To get an idea of the event characteristics due to  different types of mediators, first separate trainings were performed for the datasets of the individual scenarios. The arrangement of the mass configurations was then investigated with the goal of tracing the latent space structure back to physical properties.
The contours corresponding to a CDF value of 0.5, created with the method described in section~\ref{sec:SUSY_results}, are shown in Fig.~\ref{mediator_single_comparison}. 
To guide the eye, the coloured contours show the distributions for $m_{\rm med}=600$--2000~GeV (in steps of 200 GeV) with the dark matter mass fixed to $m_{\rm DM}=100$~GeV; the contour lines for all other mass configurations are drawn in light gray.

As can be seen, for the vector mediator data, the distributions of the individual mass configurations are heavily overlapping. 
With increasing mediator mass, the distributions get somewhat shifted and narrowed in particular in latent dimension 1, but they remain overlapping even for very different mediator masses, e.g.\ $m_{\rm med}=600$~GeV compared to $m_{\rm med}=2000$~GeV.
A similar behaviour is noticeable for the pseudoscalar mediator data. 
All in all, for both these scenarios, different mass configurations lead to overlapping representations. This is somewhat expected, since the event features are strongly dependent on the ISR jets, which have a weak dependence on the mediator mass, as illustrated by Fig.~\ref{fig:alltypes_MET_extremes}. A different behaviour is observed for the squark mediator dataset, where the hard jet spectrum is directly related to the squark-neutralino mass difference, resulting in a clear (albeit still partly overlapping) separation between different mediator masses in the latent space.

\begin{figure}[t!]\centering
\includegraphics[width=\textwidth]{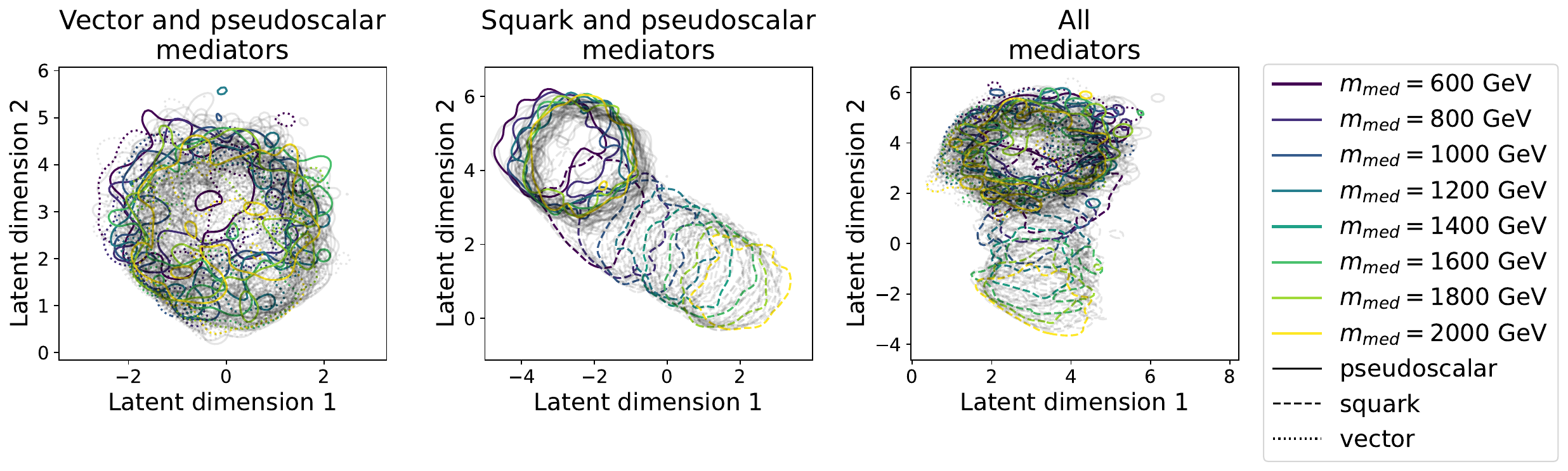}
\caption{Latent space for the embeddings of pairs of different mediator types (left and center), and for all mediator types (right). The contours correspond to a CDF value of 0.5 for the varying mediator masses, with $m_{\rm DM}=100$~GeV. The contour lines for all other mass configurations are drawn in light gray.}
\label{mediator_pair_comparison}
\end{figure}

Having seen the internal structure of the embeddings for the theories with different mediator particles, 
we next investigate how the distinct scenarios behave when embedded together in the same latent space. To this end, 
a network was trained on the combined data (including all mass configurations) from 
pairs of mediator types. 
The resulting latent spaces for a selection of mass parameters are shown in Fig.~\ref{mediator_pair_comparison}, left and center panels. For the combination of vector and pseudoscalar mediator data, no clustering with respect to the type of mediator is visible. Instead, distributions for the same mass parameters for the different mediators are approximately mapped to the same region. 
As seen in Fig.~\ref{fig:alltypes_MET_extremes}, in order to distinguish these models it would be required to select events at the tails of the MET distribution, which corresponds to a harder MET cut. Although possible, this cut would introduce a model bias which we try to avoid here. 

Once we combine the pseudoscalar and squark models, the mapping to the latent space displays a more interesting arrangement:
the network learns to embed theories with a different mediator particle in different latent space regions while maintaining their respective internal structure from Fig.~\ref{mediator_single_comparison}.
The fact that this internal arrangement of the theories remains unchanged demonstrates the stability of the embedding. 
Furthermore, we see that the squark model with $m_{\rm med} = 600$~GeV has the largest overlap with the pseudoscalar models, as is expected from Fig.~\ref{fig:alltypes_MET_extremes}. 

Finally, a training on the data from all theories was performed. The resulting latent space is shown in the right panel of Fig.~\ref{mediator_pair_comparison}. The arrangement of the theories in the latent space could be expected from the previous results. Again the vector and pseudoscalar data are mapped to the same region, close to the distributions of low-mass squark mediators. All the theories remain in the same similarity relation as observed in the pairwise embeddings. It is also interesting to notice that all the models are mostly distributed along a vertical line, while each model presents some spread around the horizontal dimension. This suggests that in this case a one dimensional latent space might be sufficient to capture the main distinguishing features of the theory space considered.
We also note that using a single vector (or pseudoscalar) benchmark model in addition to squark models with several $\Delta m$ values is sufficient to provide a good coverage of the latent space.

\begin{figure}[t!]
\centering
\includegraphics[width=0.75\textwidth]{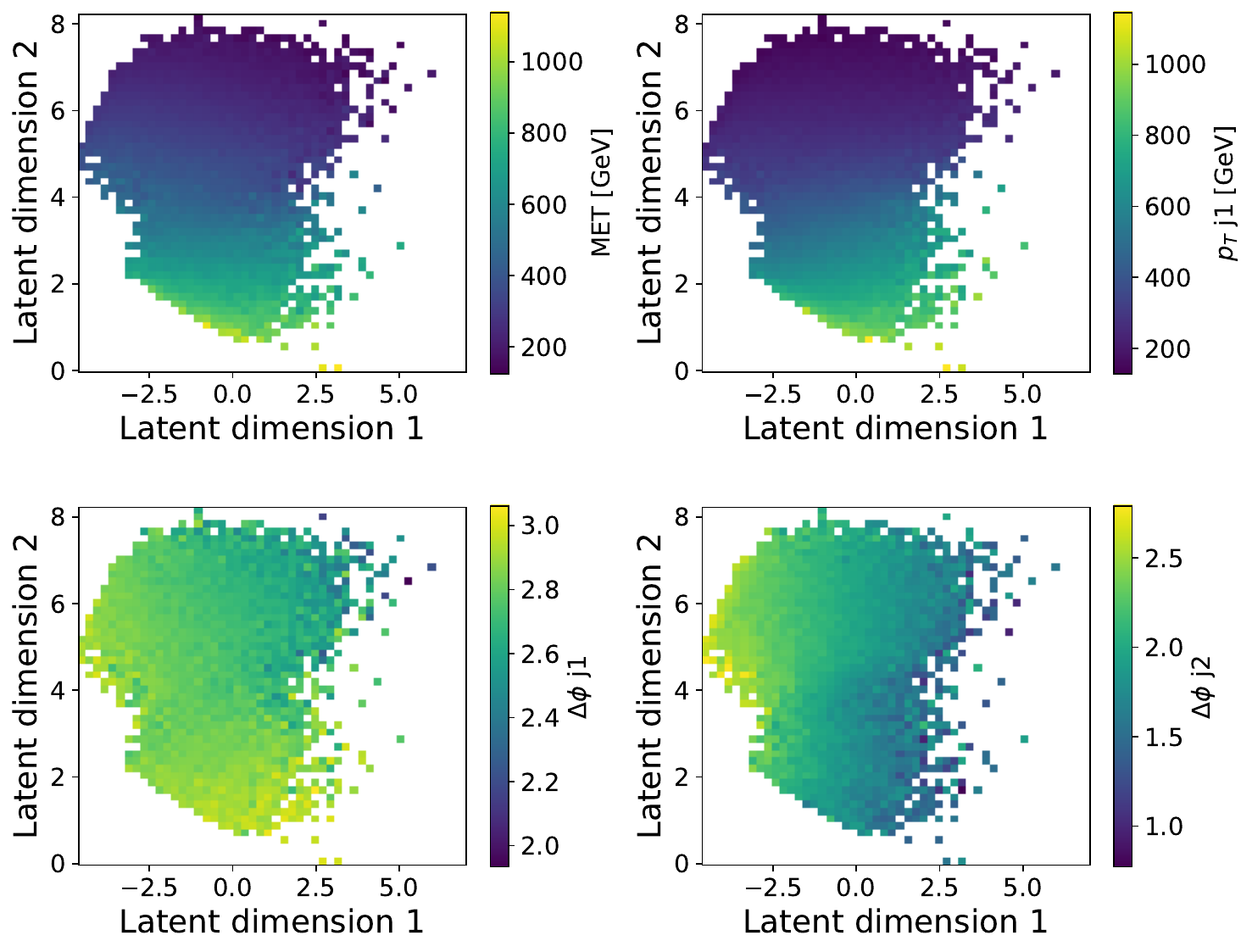}
\caption{Binned latent space after training on a combined dataset containing all mediator types and masses. The bins are colored according to the average value of the displayed input feature.}
\label{mediator_selected_input_features}
\end{figure}

To provide a better understanding of the latent space structure, 
Fig.~\ref{mediator_selected_input_features} shows the correlation of the latent space position with the MET of the event, the $p_T$ and $\Delta\phi$ of the first jet, and the $\Delta\phi$ of the second jet (see Appendix~\ref{sec:appendix_dark_matter} for the same plots for the full selection of features). Here, a binning is applied to the latent space from Fig.~\ref{mediator_pair_comparison} (right) resulting from the training on all mediator types.
For each bin, we compute the mean of the kinematic features over all data points ($\vec{X}_i$). However, since each data point carries information about a set of 10 events, the mean over the set is used to represent each data point.
We can see that the input features $p_{T,1}$ and MET both define the latent space structure along the vertical latent dimension. The fact that they have the same orientation is expected, since MET is strongly correlated with $p_{T,1}$.
The other (horizontal) latent dimension shows a strong correlation with $\Delta \phi (j_2)$. We note, however, that this feature is not very helpful when distinguishing between different models, as seen in Fig.~\ref{mediator_pair_comparison} (right).
Plots like these allow a better understanding of the learned embedding and make the structure learned by the network easier to interpret.

\section{Dark Machines Anomaly Challenge}
\subsection{Dataset}

As a final example, we consider a subset of the Dark Machines Anomaly Challenge~\cite{darkmachines_community_2020_3961917,Aarrestad:2021oeb} models. The Dark Machines data consist of simulated events for 12 different BSM models as well as a dataset with 26 different SM background processes. From these we have selected the subset of BSM models giving jets+MET signatures, shown in Table~\ref{tab:dark_machine_signals_used}. These include the gluino-neutralino scenario discussed in section~\ref{sec:susy_dataset}, the vector mediator model from section~\ref{sec:dark_matter_dataset} and two additional SUSY simplified models corresponding to stop and squark pair production with direct decays to neutralinos. We have also included the SM background dataset, which is dominated by QCD (multijets), $Z$+jets, $W$+jets and $t \bar{t}$. 

The Dark Machines data is divided into four channels, corresponding to different signal regions with different trigger requirements. 
In our analysis, we focus only on channel 1, which requires:
\begin{enumerate}
    \item at least 4 (b-)jets with $p_T > 50$ GeV,
    \item one (b-)jet with $p_T > 200$ GeV,
    \item $H_T \geq 600$ GeV,
    \item {\rm MET}$\geq 200$ GeV and ${\rm MET}/H_T \geq 0.2$,
\end{enumerate}
where $H_T$ is the scalar sum of  the transverse momenta of all (b-)jets.
The input features used for this analysis are the kinematic variables $E$, $p_T$, $\eta$ and $\phi$ of the first four jets, and the MET of the event. 
The distributions for all the features for the selected models are shown in Appendix \ref{app:DarkMachines}. 

\begin{table}[!h]
    \centering
    {\renewcommand{\arraystretch}{2}%
    \begin{tabular}{|p{5cm}|p{8cm}|p{2cm}|} 
        \hline
        BSM scenario & Physical process and model parameters  \\
        \hline
        DM Vector Mediator  & \hbox{$ p p \to Z' \to \chi \chi$}\hbox{{\textbullet} $m_{Z'} = 2$ TeV, $m_{\rm DM} = 50$ GeV}\\
        \hline
        Gluino Simplified Models  & \hbox{$p p \to \tilde g \tilde g$, $\tilde g \to q q + \tilde \chi_1^0$} \hbox{{\textbullet} $m_{\tilde{g}}=1.4$ TeV, $m_{\chi^0}=1.1$ TeV}
        \hbox{{\textbullet} $m_{\tilde{g}}=1.6$ TeV, $m_{\chi^0}= 0.8$ TeV} \\
        \hline
        Stop Simplified Model &  \hbox{$p p \to \tilde t \tilde t$, $\tilde t \to t + \tilde \chi_1^0$} \hbox{{\textbullet} $m_{\tilde{t}}=1$ TeV, $m_{\chi^0}=0.3$ TeV} \\
        \hline
        Squark Simplified Model & \hbox{$p p \to \tilde q \tilde q$, $\tilde q \to q + \tilde \chi_1^0$} \hbox{{\textbullet} $m_{\tilde{q}}=1.8$ TeV, $m_{\chi^0}=0.8$ TeV} \\
        \hline
    \end{tabular}
    }
    \caption{Dark Machines BSM models considered in our analysis. In \cite{darkmachines_community_2020_3961917,Aarrestad:2021oeb} they are called, from top to bottom: ``$Z'+$monojet'', ``SUSY gluino-gluino production'', ``SUSY stop-stop production'', and ``SUSY squark-squark production''; we have renamed them to make the correspondence with the models in sections~\ref{sec:susy_dataset} and~\ref{sec:dark_matter_dataset} apparent.}
    \label{tab:dark_machine_signals_used}
\end{table}

\begin{figure}
    \centering
    \includegraphics[width=\textwidth]{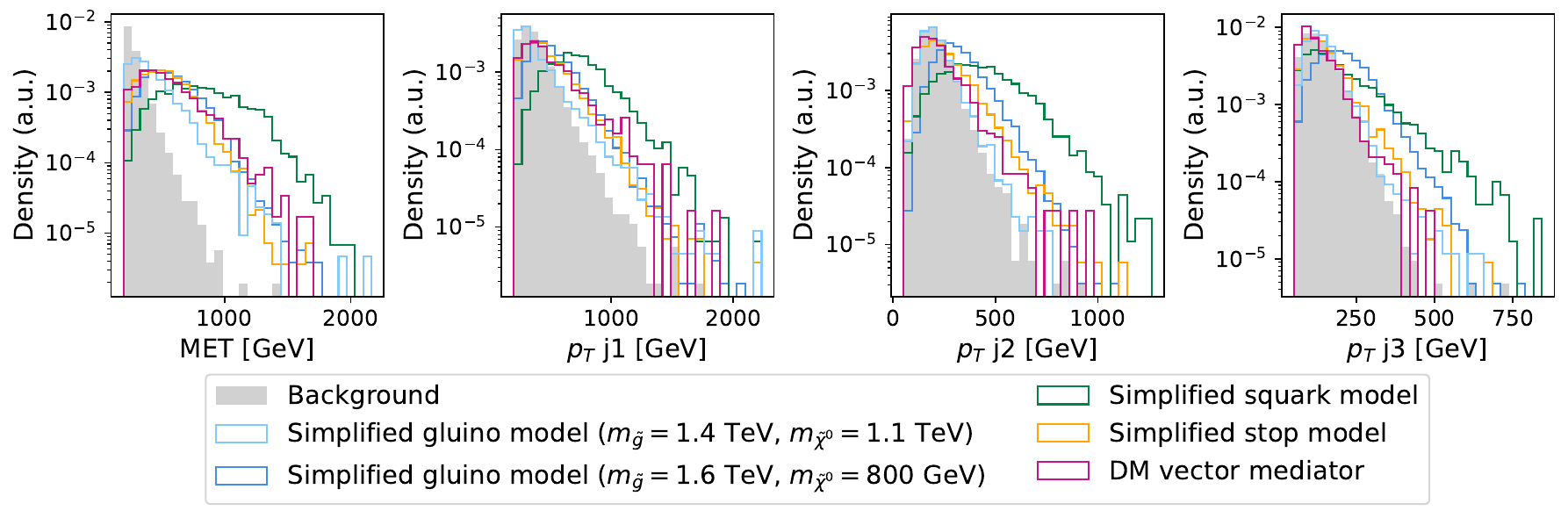}
    \caption{MET and $p_T$ distributions for the first 3 jets. The histograms for the five BSM models from Table~\ref{tab:dark_machine_signals_used} are shown in color, while the SM background corresponds to the filled gray histogram.}
    \label{fig:dists_DarkMachines}
\end{figure}

In Fig.~\ref{fig:dists_DarkMachines}, we show the MET  and the $p_T$ distributions for the first 3 jets for the selected BSM models and the SM background. The squark model has the largest mass difference ($\Delta m = 1$ TeV) and each squark decays to a single jet plus a neutralino. As a result, the MET distribution for this model is the hardest of all the models considered, as seen in the first panel of Fig.~\ref{fig:dists_DarkMachines}. All the other models have similar MET spectra, except for the light gluino scenario, which has a small mass gap ($\Delta m = 0.3$ TeV) and displays a slightly softer distribution.
A similar behavior is also seen for the first  jet $p_T$. Note that, despite having a large mass difference, all the jets in the vector mediator model come from initial state radiation and are not dependent on $\Delta m$.
Once we consider the third jet $p_T$ we see that the models start to display a different behavior. In particular, the heavy gluino model shows a peak at higher $p_T$ values when compared to all the other models.
This is expected, since each produced gluino decays to two jets and a neutralino, thus resulting in events with four hard jets. The same would also be true for the light gluino model, except for its small $\Delta m$, which leads to a softer spectrum. The other models, however, tend to display softer third and fourth jets as they originate from ISR. As a result, the stop, $Z'$ and light gluino models all display a similar $p_{T,3}$ spectrum. Finally we point out that while the SM background has very distinct (softer) MET  and $p_{T,1}$ distributions, the transverse momentum distributions of the second and third jets are similar to those of the light gluino and the $Z'$ models.

\subsection{Results}

\begin{figure}
    \centering
    \includegraphics[width=\textwidth]{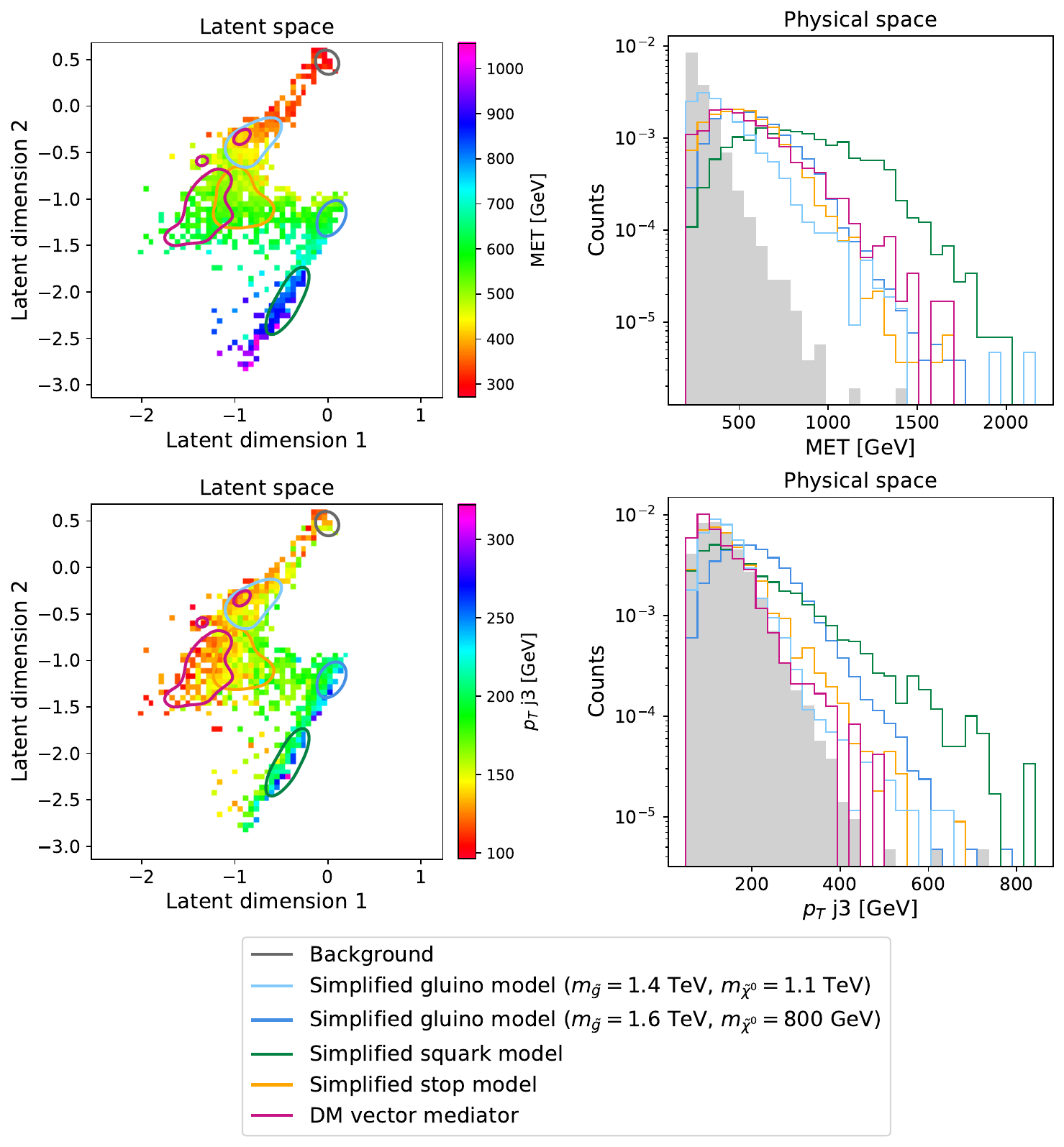}
    \caption{Left: contours of the different models in the resulting latent space overlaid on the mean of the MET and the $p_T$ of the third jet of the events placed in each particular bin in latent space. The contours  correspond to a CDF value of 0.5 for each model. Right: the distributions of the MET and the $p_T$ of the third jet for the different models.}
    \label{fig:DarkMachines_mainresult}
\end{figure}

The latent space resulting from training over all the five BSM models plus the SM background is shown in Fig.~\ref{fig:DarkMachines_mainresult}.
The left column shows the latent space contour plot (created with the same method as described in section \ref{sec:SUSY_results}) for the different signals, overlaid on the binned feature distributions of MET and $p_{T,3}$ in latent space, while the right column shows the distributions of these two quantities in physical space. These two features are highlighted here because they seem to be the main sorting axes of the features in latent space. The full binned feature distribution in latent space for all features are shown in Appendix~\ref{app:DarkMachines}.

Analyzing the contour plots, we see that the light gluino ($m_{\tilde g} = 1.4$~TeV), the $Z'$ and the stop  models all fall into a similar region of latent space. This is expected from the discussion in the previous section, where it was shown that the MET and $p_T$ distributions for these models are similar. In addition, the light gluino model displays a softer MET distribution, which puts this model at slightly higher values for the second (vertical) latent space dimension. The background has the softest MET distribution, thus being located at the very top of the latent space plot.
The squark and the heavy gluino models, on the other hand, populate disjoint regions. While the squark model is located at the bottom of the plot, which corresponds to the highest MET values, the heavy gluino model occupies the center-rightmost region. This region corresponds to intermediate MET values and large $p_{T,3}$ values, as expected from the previous discussion.

In summary, the network was able to explore the signal features of each model and cluster them accordingly. We also point out that the $Z'$, light gluino and stop models have similar features, which means that for benchmarking purposes just one of them is sufficient to represent this region in latent space.  In turn, an additional benchmark model with a similar MET but a harder $p_T(j_3)$ spectrum would result in a better coverage of the signature space.

\section{Conclusions}

Motivated by the large phase space of BSM models, we proposed a strategy to build a joint embedding space of theories.
The main property of this abstract space is that models are clustered according to their phenomenology (i.e., their experimental signatures), 
so different models are mapped to different regions in this space while indistinguishable models are mapped to the same region.

We demonstrated the feasibility of this technique in three examples of increasing complexity. First, we considered a simplified MSSM scenario 
with different gluino and neutralino masses and observed that the embedding without additional guidance correctly identifies the gluino to neutralino mass difference as the phenomenologically most relevant parameter and arranges the models by this quantity in latent space accordingly.

Next, we investigated three different models of dark matter production in final states with a hard jet and missing energy: a vector mediator, a pseudoscalar mediator, and a squark mediator model. We observed that the learned representations for the vector mediator and pseudoscalar mediator do not vary significantly with the mediator mass, mostly because the jets of these events come from ISR. 
For the squark mediator, on the other hand, where the mass of the jet is directly related to the mass difference between the squark and neutralino, different masses are embedded in different regions of latent space. When learning a joint embedding of all three models, the same picture remains: vector and pseudoscalar mediator are mapped to one region, which is phenomenologically indistinguishable from a low mediator mass squark model. Higher mediator mass squark models extend away from this point, ordered by their mass.

The third showcase was taken from the Dark Machines Anomaly Challenge dataset. Here, we observed a sorting of the latent space along the direction of missing transverse energy and the $p_T$ of the third jet. The network's focus on MET was to be expected, given how distinguished the MET distributions of the background and the squark model are compared to the remaining models. In addition, the network was able to separate models with similar MET spectra by exploiting the $p_T$ of the third jet, which puts the heavy gluino model far from the stop and DM vector mediator models. Therefore, the latent space can not only cluster together models with similar features, but also indicate possible discriminating observables.

The three examples investigated here have illustrated that the proposed approach can achieve the clustering of models with similar features, thus providing a way to select a minimal set of benchmark models with a maximal coverage of the feature space.
In addition, the correlation between the latent space dimensions and the physical features can be used to identify suitable observables for experimentally distinguishing BSM models.
Finally, regions in latent space not covered by any of the chosen models can indicate gaps in signature space, which could be filled by different benchmark choices, thus maximizing the physical signatures explored at the LHC and future colliders.

Together, these results demonstrate the potential usefulness of a learned representation to combine and understand data from different theoretical models. 
However, to fully realize the potential of this approach, more work is needed. Most important is an inclusion of the cross-section of a given process in addition to the distributions of features, as this is an important property that would be useful for separating different models. Next a more expressive basis of features and higher capacity learning algorithms need to be explored. 
It will also be useful to develop an algorithm that is capable of sampling the latent space and reconstruct the physical features corresponding to these sampled points. This would allow us to explore unpopulated regions of latent space, which could potentially represent signal features which are not present in any of the BSM models considered. 
Together this would motivate the building of a universal embedding of BSM theories to guide the exploration of model space at the LHC and beyond.

\section*{Acknowledgements} 
This research was supported in part through the Maxwell computational resources operated at Deutsches Elektronen-Synchrotron DESY, Hamburg, Germany. The work of AH, GK, LM and TS was supported by the Deutsche Forschungsgemeinschaft under Germany’s Excellence Strategy – EXC 2121  Quantum Universe – 390833306, and under PUNCH4NFDI – project number 460248186. The work of SK was supported in part by the IN2P3 theory master project ``DataMATTER''. AL~is supported by FAPESP grants no.~2018/25225-9 and 2021/01089-1\@. 
DS is supported by the U.S. Department of Energy (DOE), Office of Science grant DOE-SC0010008 

We thank the organizers of the 2022 MITP workshop A Deep-Learning Era for Particle Theory, the Les Houches PhysTeV 2023 BSM workshop, and the Aspen Center for Physics (supported by National Science Foundation grant PHY-2210452) for opportunities to discuss this work.

\clearpage

\appendix
\section{Supplementary material} 

\subsection{MSSM gluino dataset}

\label{app:MSSM}

\begin{figure}[h!]
    \centering
    \includegraphics[width=0.95\textwidth]{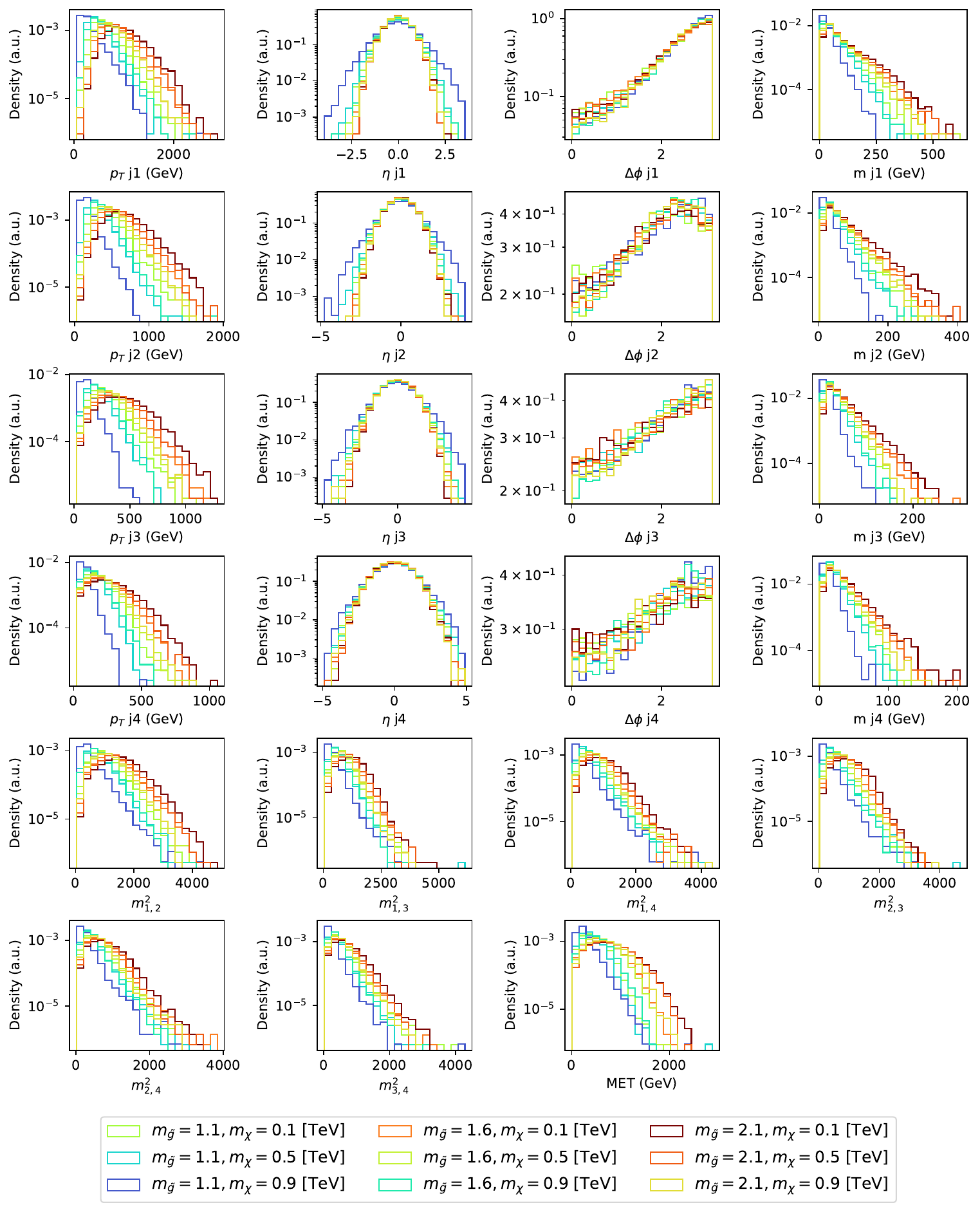}
    \caption{Feature distributions of the MSSM gluino dataset.}
    \label{fig:mssm_allfeatures}
\end{figure}

Figure \ref{fig:mssm_allfeatures} shows the distributions of the features used in the training on the MSSM gluino dataset.

\clearpage

\subsection{Dark Matter dataset}
\label{sec:appendix_dark_matter}

\begin{figure}[h!]
    \centering
    \includegraphics[width=0.95\textwidth]{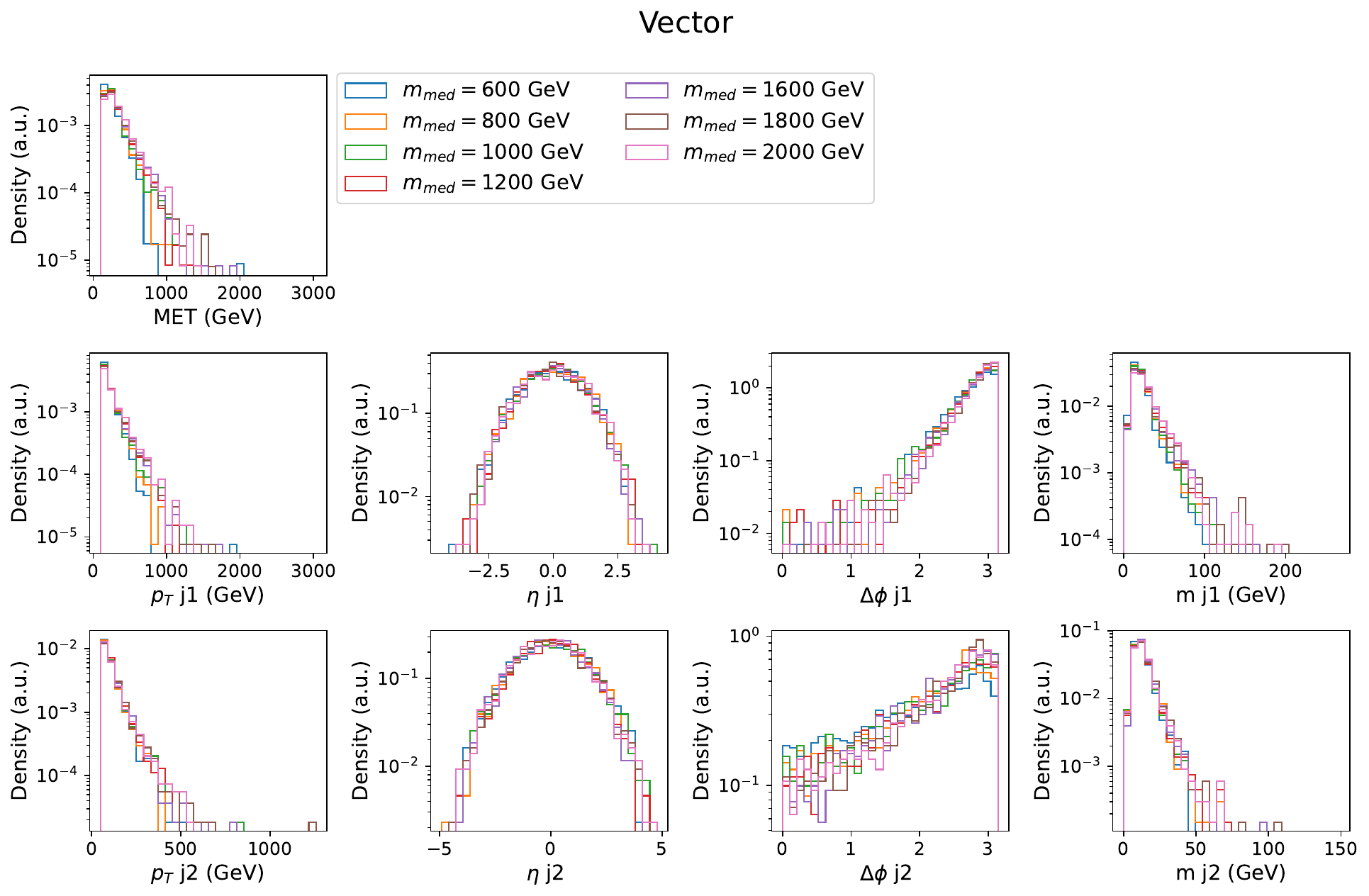}
    \caption{Features of the vector mediator scenario for the different mediator masses with $m_{\rm DM}=100$~GeV.}
    \label{fig:vector_features}
\end{figure}

\begin{figure}
    \centering
    \includegraphics[width=0.95\textwidth]{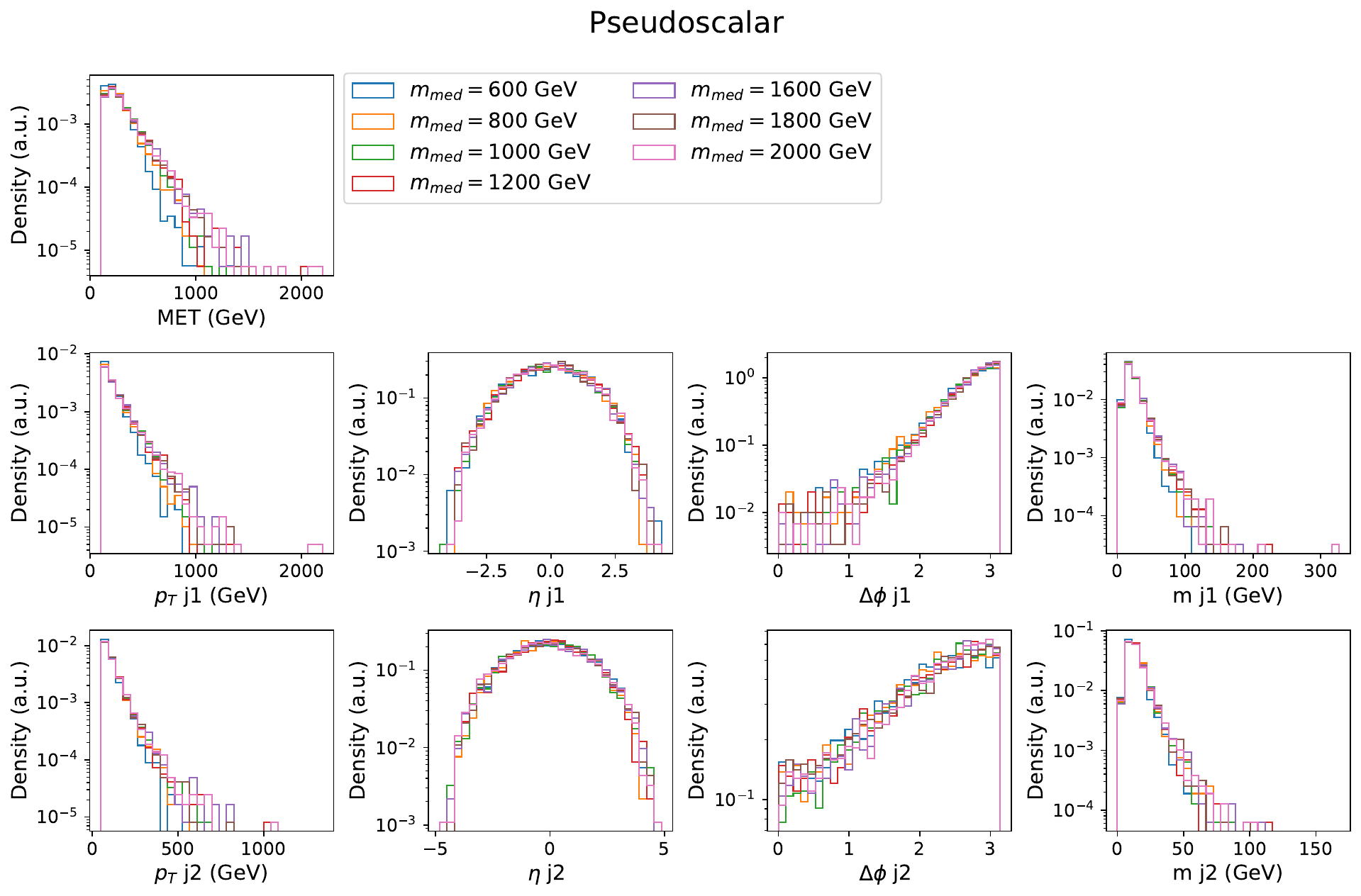}
    \caption{Features of the pseudoscalar mediator scenario for the different mediator masses with $m_{\rm DM}=100$~GeV.}
    \label{fig:pseudoscalar_features}
\end{figure}

\begin{figure}
    \centering
    \includegraphics[width=0.95\textwidth]{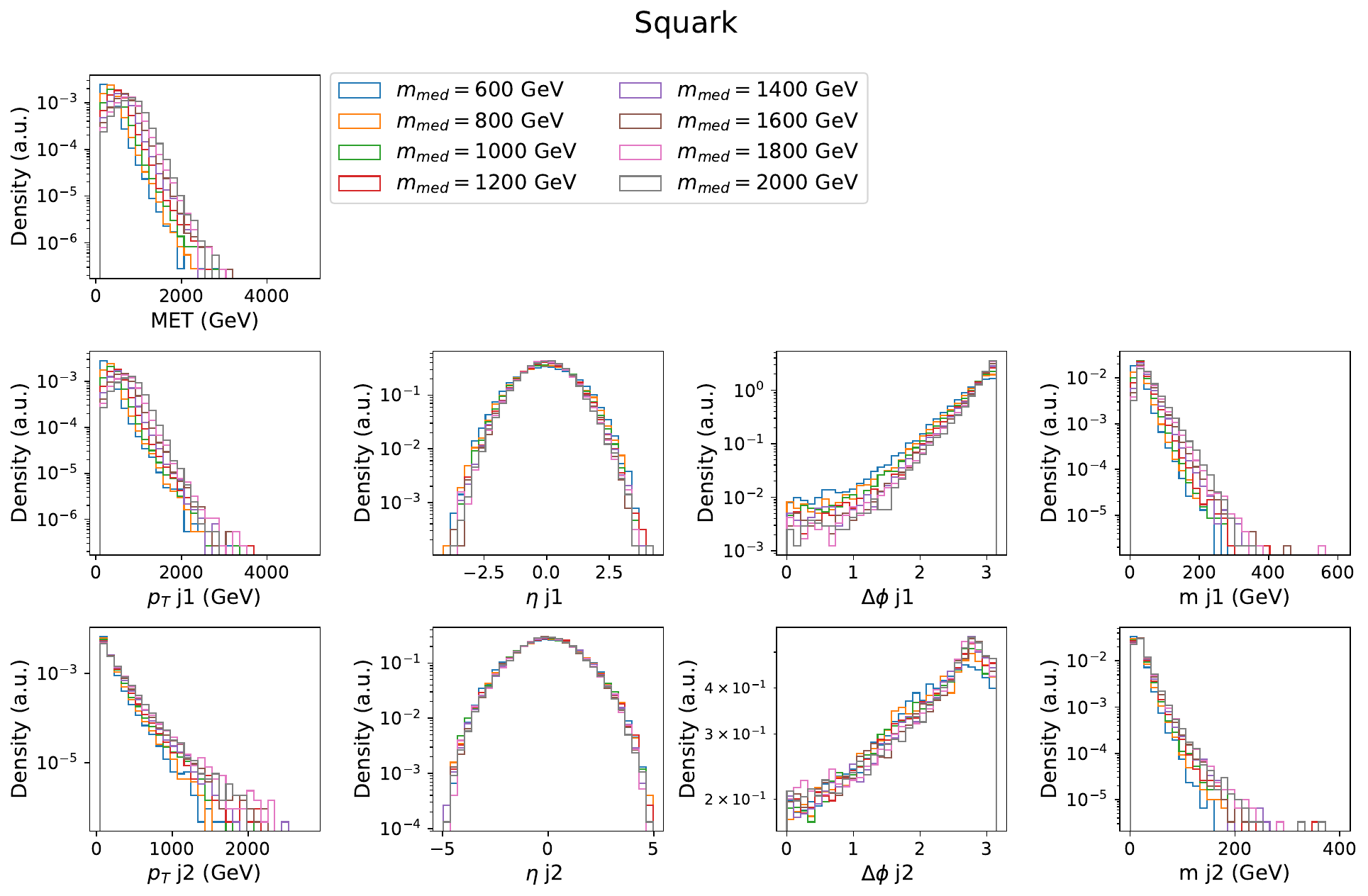}
    \caption{Features of the squark mediator scenario for the different mediator masses with $m_{\rm DM}=100$~GeV.}
    \label{fig:squark_features}
\end{figure}

\begin{figure}[t]
\centering
\includegraphics[width=0.75\textwidth]{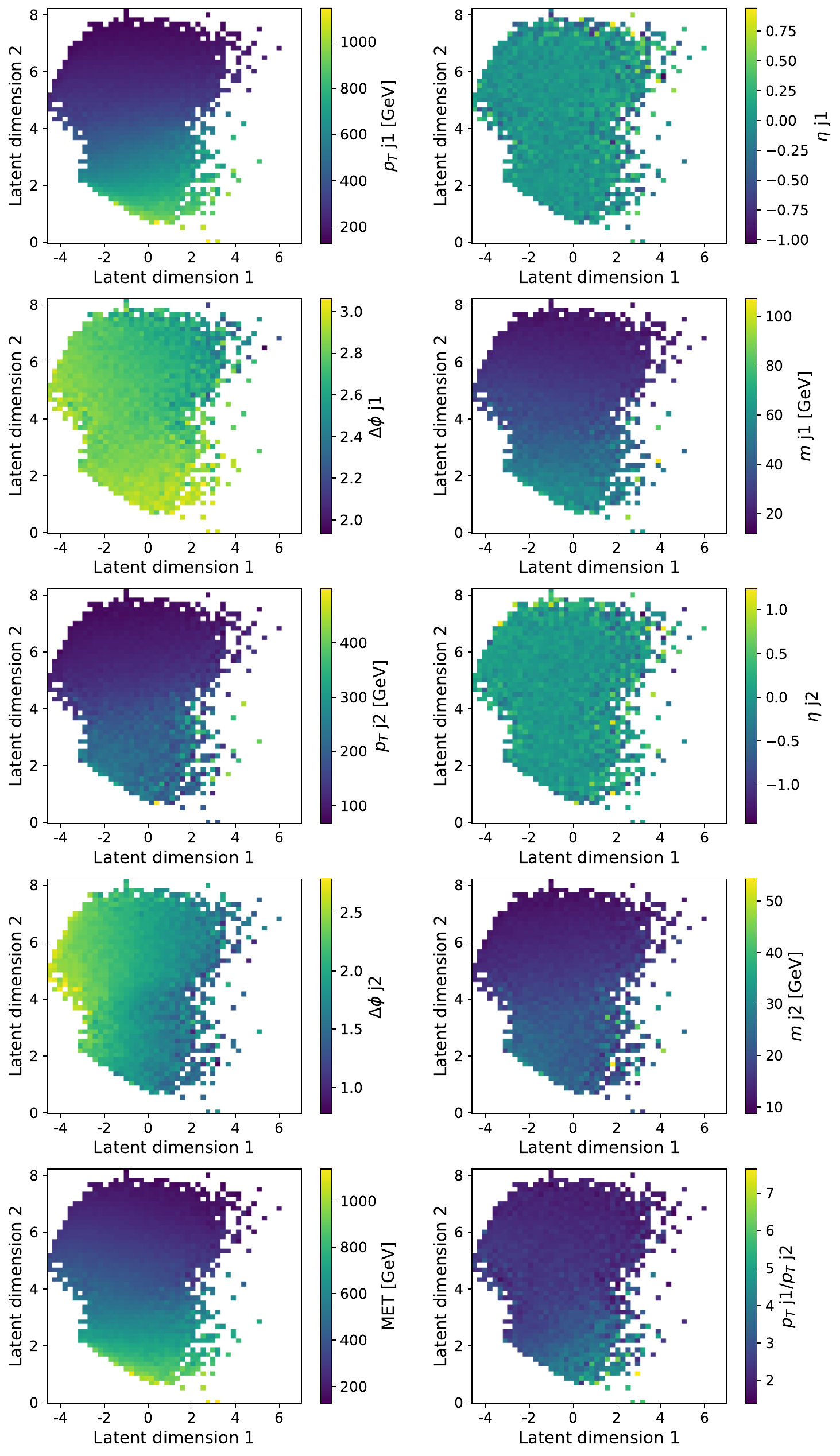}
\caption{Binned latent space after training on a combined dataset containing all mediator types and all mass configurations in the Dark Matter dataset.}
\label{mediator_all_input_features}
\end{figure}

Figures~\ref{fig:vector_features}, \ref{fig:pseudoscalar_features} and \ref{fig:squark_features} show the feature distributions for the three different mediator scenarios, for the different mediator masses and $m_{\rm DM}=100$~GeV. 
Figure~\ref{mediator_all_input_features} shows the binned latent space for the different features after training on a combined dataset of all mediator types of all mass combinations. 

\clearpage

\subsection{Dark Machines dataset}
\label{app:DarkMachines}
\begin{figure}[!h]
    \centering
    \includegraphics[width=\textwidth]{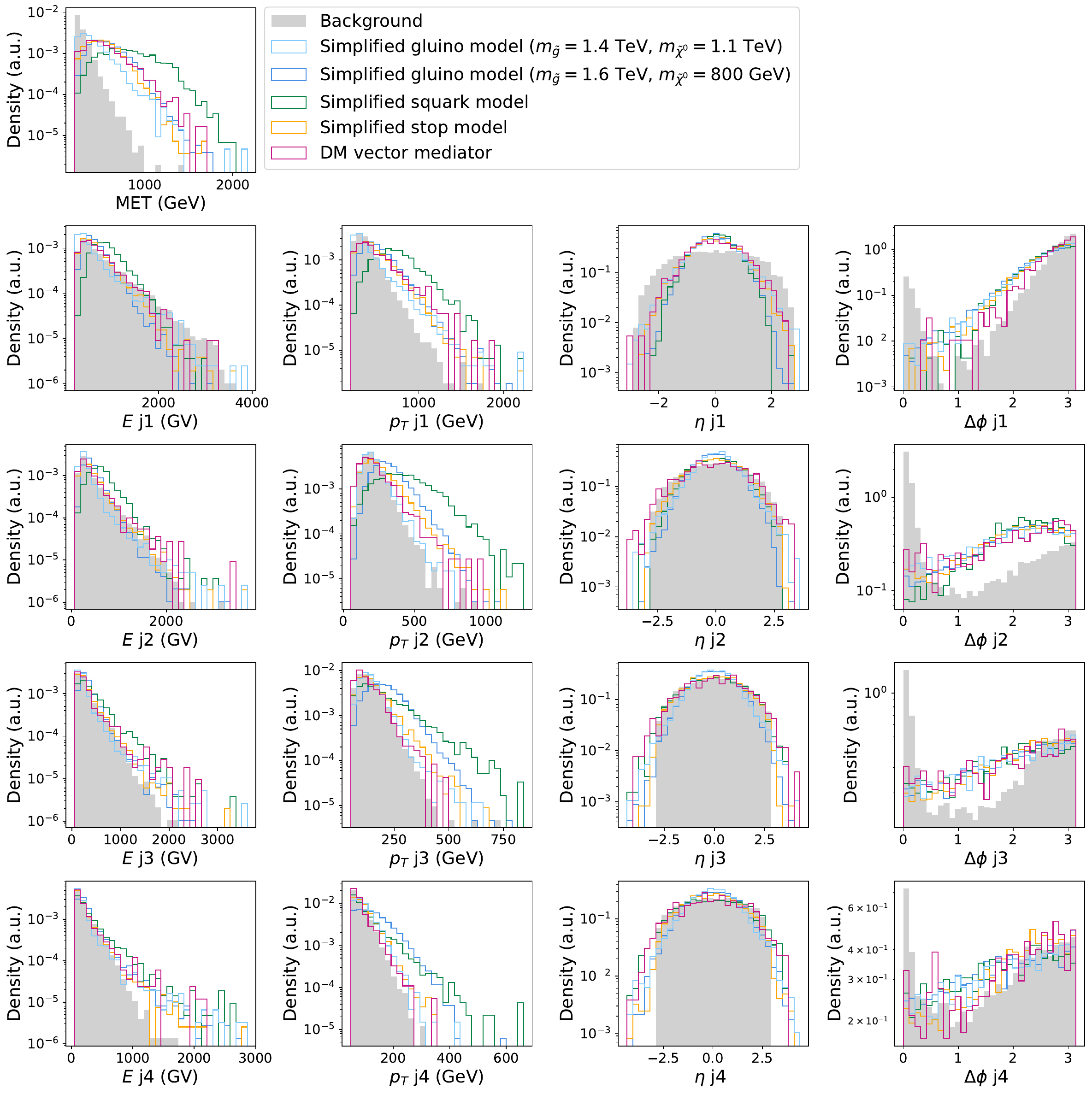}
    \caption{Feature distributions for the selected signals from the Dark Machines dataset.
    }
\label{fig:DarkMachines_histo_ch1_only}
\end{figure}

\begin{figure}[!h]
    \centering
    \includegraphics[width=\textwidth]{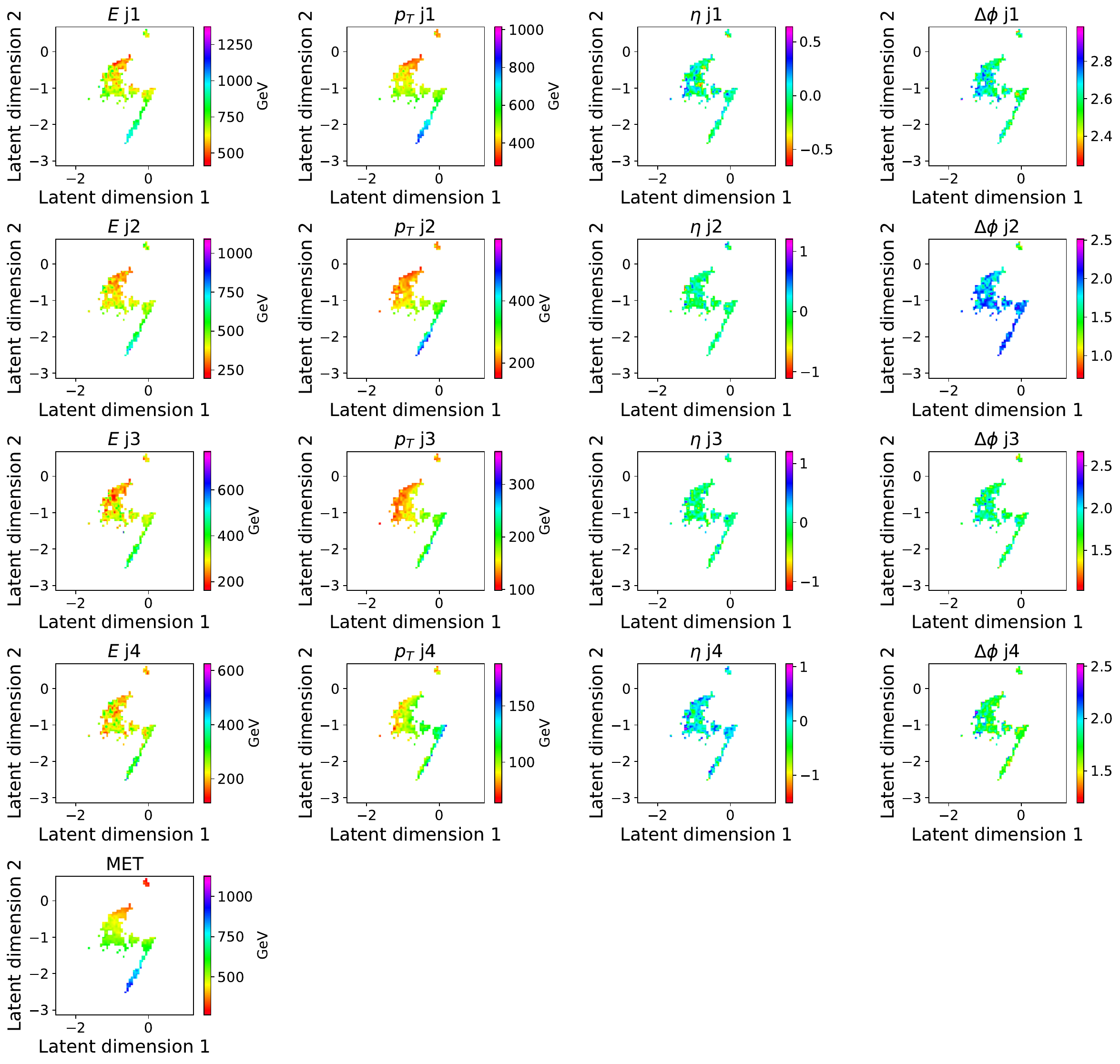}
    \caption{Binned feature distributions in the latent space of the Dark Machines study.}
    \label{fig:DarkMachines_binned_ch1_only}
\end{figure}

Figure~\ref{fig:DarkMachines_histo_ch1_only} shows the distributions of the different features used for training in the Dark Machines dataset. Figure~\ref{fig:DarkMachines_binned_ch1_only} shows the binned feature distributions in the latent space of the Dark Machines study.

\clearpage

\bibliography{bibliography}

\end{document}